\documentclass[useAMS,usenatbib]{mn2e}
\usepackage{epsfig}

\title[Long $\Gamma$-ray Bursts from Common Envelope Evolution]
{A Common Envelope Binary Star Origin of Long Gamma-ray Bursts}

\author[C. A. Tout, D. T. Wickramasinghe, H. H.-B. Lau, J. E. Pringle \& L. Ferrario]
{Christopher A. Tout$^{1,2,3}$, Dayal T. Wickramasinghe$^2$, Herbert H.-B. Lau$^3$,
\newauthor
J. E. Pringle$^1$ and Lilia Ferrario$^2$\\
$^1$Institute of Astronomy, The Observatories, Madingley Road,
Cambridge CB3 0HA\\
$^2$Mathematical Sciences Institute, The Australian National University,
ACT 0200, Australia\\
$^3$Centre for Stellar and Planetary Astrophysics, Monash University,
PO Box 28M, Victoria 3800, Australia}
\begin{document}
\date{Accepted.  Received ; in original form} 
\pagerange{\pageref{firstpage}--\pageref{lastpage}} \pubyear{}

\maketitle

\label{firstpage}

\begin{abstract}
  The stellar origin of $\gamma$-ray bursts can be explained by the
  rapid release of energy in a highly collimated, extremely
  relativistic jet.  This in turn appears to require a rapidly
  spinning highly magnetised stellar core that collapses into a
  magnetic neutron star or a black hole within a relatively massive
  envelope.  They appear to be associated with type~Ib/c supernovae
  but, with a birthrate of around $10^{-6}-10^{-5}$\,per year per
  galaxy, they are considerably rarer than such supernovae in
  general. To satisfy all these requirements we hypothesize a binary
  star model that ends with the merging of an oxygen neon white dwarf
  with the carbon-oxygen core of a naked helium star during a
  common envelope phase of evolution.  The rapid spin and high
  magnetic field are natural consequences of such a merging.  The
  evolution that leads to these progenitors is convoluted and so
  naturally occurs only very rarely.  To test the hypothesis we evolve
  a population of progenitors and find that the rate is as required.
  At low metallicity we calculate that a similar fraction of stars
  evolve to this point and so would expect the $\gamma$-ray burst rate
  to correlate with the star formation rate in any galaxy.  This too
  is consistent with observations.  These progenitors, being of
  intermediate mass, differ radically from the usually postulated
  high-mass stars.  Thus we can reconcile observations that the bursts
  occur close to but not within massive star associations.
\end{abstract}


\section {Introduction}
There is now strong evidence that the long $\gamma$-ray bursts (LGRBs)
are closely associated with young star forming regions
\citep{woosley2006a}.  Recent high spatial resolution imaging of
nearby LGRBs has shown that, while they are closely associated with
clusters of Wolf-Rayet stars and of O stars, they tend also to be
displaced somewhat from the centres of the clusters
\citep{hammer2006}.  This gives further insights into their population
characteristics.

The birth rate of LGRBs is estimated to be $10^{-6} - 10^{-5}\,\rm
yr^{-1}$ per galaxy like our own when allowance is made for the
uncertainties in the beaming angle of the $\gamma$-rays.  This is
$1\,000 - 10\,000$ times lower than the birth rate of type~II
supernovae \citep{fryer2007}.  The LGRBs in which optical
afterglows have been detected tend in general to be about a $100$
times more luminous than type II SNe and this has led to the
suggestion that they form a new class of supernovae or hypernovae
\citep {paczynski1998, woosley1993}.  The association with young star
clusters also places them in a class distinct from type Ia SNe which
are thermonuclear explosions of degenerate white dwarfs.  At the same
time, there have been unambiguous identifications of a few LGRBs with
type~Ib/c supernovae which are characterised by hydrogen- or hydrogen-
and helium-deficient ejecta.

Type~Ib/c supernovae are expected at the ends of the lives of massive
stars that have lost much of their mass during Wolf-Rayet evolution so
the evidence suggests that the LGRBs are linked in some way to
the final stages of the evolution of very massive stellar cores as
they collapse to black holes or neutron stars.  The total energy, of
about $10^{51}\,$erg, released is not much greater than that released
in core collapse supernovae.  However it differs in that the energy is
released in $\gamma$-rays in the form of collimated jets.  These
properties have led to the suggestion that the LGRBs form a subset of
core collapse supernovae which are distinguished by the magnetic
fields present in the core that collapses to a black hole or a neutron
star.  These fields collimate the jet.  In a popular model, the
$\gamma$-ray jet is believed to be launched by the magnetohydrodynamic
(MHD) extraction of the spin energy of a disrupted torus or a central
rapidly spinning black hole \citep[the collapsar
  model,][]{meszaros1997}.

The Lorentz factor $\Gamma$ for $\gamma$-ray burst jets is about
400 \citep{lithwick2001}. This is in stark contrast to the
velocities of other observed astrophysical jets. For non-relativistic
jets, the jet velocities are typically only a few times the escape
speeds from the central accreting objects \citep{price2003}.  For jets
from active galactic nuclei (AGN), where the central object is a
supermassive black hole, jet velocities typically have Lorentz factors
in the range $3 \le\Gamma\le 10$ \citep{giovannini2001} and
occasionally slightly higher, perhaps as high as $\Gamma \approx 20$
\citep{giroletti2004,hough2008}. Thus it is evident that something
special must be going on in the engines which produce GRBs which is
not occurring in the engines which produce most other astrophysical
jets. To produce an astrophysical jet it is evident that a large
amount of the available accretion energy must be given to a small
fraction of the material and such a process is most easily
accomplished by making use of magnetic fields. From numerical and
analytic models it is also evident that the production of a strong jet
requires the presence of strong rotation (usually in the form of an
accretion disc flow) and a strong {\it poloidal} magnetic field
\citep[see for
  example,][]{tchekhovskoy2008,lyubarsky2009,komissarov2009}.  A more
complicated field structure tends to weaken jet production
\citep{beckwith2008}.

In this regard the collapsar model
\citep[e.g.][]{macfadyen1999,woosley2006a} which involves the collapse
of a strongly magnetic, rotating stellar core is the most
promising.  Rotation ensures that the collapse is slowed, so that for
the case of collapse to a black hole maximum energy extraction is
possible, and an accretion disc forms, so that the geometry is conducive to jet formation. In
addition, because the accretion rate is so high (compared for example
to AGN discs), being in the range of $0.1 \le \dot{M}/{\rm
  M_\odot\,s^{-1}} \le 10$ the disc is geometrically thick (thickness
$H$ is comparable to radius $R$) and advection dominated
\citep{dimatteo2002}. This means that, in contrast to AGN discs which
are geometrically thin at least in their outer regions and so require
field generation by local dynamo action, a poloidal magnetic field
already present in the infalling material can be dragged inwards
\citep{lubow1994} and thus compressed and strengthened. In the centre
of such a flow a compact object is likely to form either a magnetar, a
highly magnetic neutron star or a black hole.  Subsequent to the
formation it is then possible to extract the rotational energy of such
objects, for black holes through the Blandford-Znajek mechanism
\citep[e.g.][ for discussion]{komissarov2009,livio1999} and for
neutron stars as magnetars.

Thus it is possible that a strongly magnetic, rapidly spinning neutron
star \citep[the magnetar model,][]{usov1992,kluzniak1998,spruit1999}
accelerates and collimates the jets.  This model requires as essential
ingredients magnetic fields of about $10^{15}\,$G and rotation periods
of a few milliseconds to explain the observed release of some
$10^{52}\,$erg on the time scales of about 10\,s which are
characteristic of the LGRBs.  Whether such end products result from
cores that collapse in the course of merging with another star
\citep{woosley1993,paczynski1998,macfadyen1999} or whether they are
the result of peculiar stellar evolution to an anomalously rapid
rotation of a progenitor star as a consequence of low metallicity
\citep{woosley2006,yoon2005} remains an open question.

The need for a binary companion, particularly to generate the rapid
spin of the collapsed core immediately after formation, was pointed out
by \citet{izzard2004} but has rather been put to one side in favour of
rapidly spinning low-metallicity massive single stars.  However
\citet{podsiadlowski2004}, noting the link with hypernovae, reinforced
the need for a binary companion.  \citet{fryer2005} argued that the
collapsar model requires two massive stars of very similar mass.

Here we propose a new binary scenario that differs radically in that
the stars are of intermediate mass.  As single stars these would not
undergo a supernova explosion at all but end their lives as
carbon/oxygen or oxygen/neon white dwarfs. It is their duplicity, so
that the two cores can be merged, which allows the supernova explosion
to take place.
This model differs from the exposed accretion induced collapse
discussed by \citet{yi1997} because the collapse follows the merging of two
cores within a common envelope phase of evolution.  This has two
important consequences.  First there is mounting evidence that white
dwarfs which merge during common envelope evolution develop the strongest
magnetic fields found in white dwarfs \citep{tout2008} so that these
collapsing cores are highly magnetic.  Secondly the
remaining envelope, which must be hydrogen free in this case, allows
for the associated Ib/c supernova.

\section{Problems with current massive star models}

One possibility, that has been aired, is that the LGRB phenomenon only
occurs in the small subset of massive stars that are born rapidly
rotating and hence evolve to produce rapidly rotating remnants.
However, a rapidly spinning pre-collapse stellar core can also be
expected to generate a strong magnetic field by a dynamo
mechanism. This would brake its rotation by the transfer of angular
momentum to the outer envelope that is mostly lost during the
Wolf-Rayet phase prior to its collapse to a neutron star or a black
hole.  Thus, if strong fields are required, the end product is likely
to be a slowly rotating stellar remnant, whether it be a neutron star
or a black hole.  There is also the question of whether fields that
are strong enough to yield the very high-field ($10^{15}\,$G) neutron
stars when compressed can be generated in a pre-collapse core during
the evolution of a single star \citep{duncan1992}.

The above conclusions are borne out by detailed stellar evolution
calculations of medium and high mass stars that lead to the formation
of neutron stars and black holes.  These calculations, which allow for
the generation of small scale magnetic fields by various
instabilities, such as the magneto-rotational \citep{spruit2002}, in
differentially rotating radiative regions but not for dynamo generated
fields in convective regions, show that angular momentum is
effectively transported away from the core by magnetic stresses
\citep{heger2005}.  The end product is typically a slowly rotating
neutron star with a birth magnetic field of $10^{12}$\,G and spin
period of $0.1\,$s similar to a normal pulsar.
 
Single stars do not generally spin fast enough, so that the postulated
rapid spin is likely to have involved a binary interaction
\citep{izzard2004}.  Another possibility that has been considered is
that the progenitor star is both rapidly rotating and metal deficient.
At low metallicity wind mass loss is suppressed and this leads to
enhanced mixing so that the red-giant phase is not as pronounced.
Strongly magnetic and rapidly spinning end products then become a
possibility \citep{woosley2006}.  However, although observations do
show that LGRBs tend to occur preferentially in low-metallicity
galaxies, they do not exclusively occur in such galaxies
\citep{sollerman2005}.  Indeed \citet{savaglio2010} finds that the
$\gamma$-ray burst rate in a given galaxy simply reflects the star
formation rate in that galaxy.  So a more general mechanism which is
not entirely dependent on metal abundance appears to be
required. Moreover, there remains the problem that any single star
model is likely to overproduce LGRBs unless they can be restricted to
a very small progenitor mass range.

Evidence that LGRBs tend to be displaced by $400 - 800\,$pc from star
forming regions \citep{hammer2006} also poses problems for single star
models. In an attempt to explain the above observations it has been
proposed that an even more massive companion was able to spin up the
progenitor before exploding itself and ejecting the rapidly spinning
Wolf-Rayet star at high velocity from its birthplace.  Even if the
lone Wolf-Rayet star does not spin down because its wind is weak
at its low metallicity, we might ask why exactly the first
star, which would also have been spun up, would not have generated a
burst itself.

Those invoking binary star models have also concentrated on massive
progenitors.  \citet{fryer2005} examine merging helium cores during a
common envelope phase.  These have the advantage of generating high
spin rates in the collapsing cores and so are very promising.  However
they require two very massive stars of similar mass and the evolution
to the point of burst is very rapid.  They would therefore be expected
to occur close to the centroids of star forming regions which appears
to be inconsistent with observations.  Alternatively
\citet{cantiello2007} invoke accretion from a more massive companion
as necessary to spin up the Wolf-Rayet star while it is still on the
main sequence.  They argue that it does not subsequently tidally spin
down as the orbit widens or as it loses mass.  They further use a
supernova explosion and kick in the originally more massive companion
to account for the apparent runaway nature of the Wolf-Rayet
progenitor.
 
Thus it would seem that the general consensus is that the immediate
progenitor of a $\gamma$-ray burst must be compact, either a neutron
star or black hole, rapidly spinning, with a period of less than a few
milliseconds and be associated with a very strong, $10^{15}\,$G,
magnetic field.  The association with type~Ib/c supernovae further
requires that the progenitor be at the core of a naked helium or even
more processed envelope.

\section{An Intermediate-Mass Binary Star Scenario}

In order to form close binary stars with compact components it is
necessary to pass through a common envelope phase of evolution
\citep{paczynski1976}.  Unstable mass transfer from a giant star that
expands as it loses mass to a more compact star leads to a common
giant-like envelope around two cores, the giant's own degenerate core
and the compact companion.  Friction between the orbit of the cores
and the envelope causes the cores to spiral together and the envelope
to unbind.  \citet{tout2008} have demonstrated that the highest
magnetic fields, of more than $10^9\,$G, are most likely generated in
common envelope evolution when the two cores merge before the envelope
is ejected.  In the scenario which we propose here, the carbon/oxygen
(CO) core of a naked helium giant merges with an oxygen/neon (ONe)
white dwarf in a final helium common envelope.  The CO core acquires a
very high magnetic field from the common envelope, is tidally broken
up and accretes on to the ONe white dwarf carrying the magnetic field
with it.  On reaching the Chandrasekhar mass the ONe white dwarf
undergoes accretion induced collapse to a neutron star.  By conserving
its magnetic flux it acquires a large scale surface field of
$10^{15}\,$G.  Both the contraction and the accretion of high angular
momentum material ensure that it is rapidly spinning and the
conditions are ripe to launch the relativistic jets required to drive
the $\gamma$-ray burst.

However the volume of parameter space which can give rise to this GRB
progenitor is not large and this can account for the scarcity of
LGRBs.  A typical system that leads to such a progenitor begins life
as a relatively wide binary system with a 6~and an $8\,\rm M_\odot$
star, both on the main sequence, and an orbital separation of
$1000\,\rm R_\odot$.  The $8\,\rm M_\odot$ star evolves through
hydrogen, helium and carbon core burning to a super-AGB star with an
ONe core and then fills its Roche lobe for the first time.  A first
phase of common envelope evolution ensues and results in a mild
shrinkage of the orbit and loss of the envelope to leave a massive ONe
white dwarf, of about $1.4\,\rm M_\odot$ with the $6\,\rm M_\odot$
companion at $160\,\rm R_\odot$.  Subsequently the $6\,\rm M_\odot$
star evolves to the early red giant branch and fills its Roche lobe.
A second common envelope phase removes its hydrogen envelope to leave
a $1.3\,\rm M_\odot$ naked helium star in a close orbit with the ONe
white dwarf at $1.4\,\rm R_\odot$.  The naked helium star develops a
CO core, evolves to a giant and fills its Roche lobe for the second
time.  The third and final common envelope phase builds up the very
strong magnetic field and causes the CO core to merge with the ONe
white dwarf which collapses to a rapidly spinning neutron star and
launches the relativistic jet in the process.
The remainder of the CO core accretes on to the neutron star at such a
high rate that carbon ignites and runaway thermonuclear reactions
generate the $^{56}$Ni in strong winds that can drive off any
remaining helium envelope and power the type Ib/c supernova
\citep{woosley2006a}.

The central engine of our GRB is the accretion induced collapse of a
highly magnetic, rapidly spinning white dwarf and so is similar to
that discussed by \citet{yi1997}.  \citet{fryer1999} modelled a
non-magnetic collapse in two dimensions and deduced that jets would be
too weak, because there would be too much material in the jet, unless
highly beamed.  \citet{dessart2007} included magnetohydrodynamics in
similar two-dimensional collapse calculations and in one model, with a
large initial magnetic field, produced a jet with enough energy to
power a $\gamma$-ray burst but still with too much material in the jet
to be accelerated to the required velocity.  This is often called the
baryon loading problem.  Other numerical simulations, even those that
examine accretion on to black holes such as those by
\citet{porth2010}, have similar problems However jet production and
collimation is still far from fully understood
\citep{lyubarsky2010,komissarov2010} and so we regard these
models as promising rather than as creating an insurmountable problem
with this scenario.

\section{Estimated Gamma-ray Burst Rates}

To estimate the rate at which such systems would give rise to LGRBs we
have carried out binary population synthesis with the code developed
by \citet{hurley2002}.  Their standard prescription for common
envelope evolution is included.  Their $\alpha_{\rm CE}$ parameter,
the efficiency of transferring orbital energy to the envelope during
common envelope evolution, is set to~1.  Though this parameter is very
uncertain, we do not investigate its effects in detail because the
observed $\gamma$-ray burst rate is even more uncertain.

At solar metallicity, $Z = 0.02$, the range of possible initial
separations that lead to the described systems is narrow.  Their
initial separations are mostly around $1000\pm 25\,\rm R_\odot$.  The
precise range depends on the component masses of the system.  If the
system is too wide then either the third common envelope phase or the
merging event is avoided.  If the system is too close the ONe white
dwarf accretes enough material to collapse to a neutron star before
the common envelope forms.  The actual distribution of initial periods
of binary stars is not well known.  A common practice is to assume the
separation is uniform in logarithmic space \citep{eggleton1989}.
With this assumption, only about $5-6\times10^{-3}$ of systems have
suitable initial separations.  A second requirement is that one
component must be massive enough to develop an ONe core.  To do so its
core must ignite carbon gently before reaching the Chandrasekhar limit
and become a super-asymptotic giant branch (SAGB) star.  The mass
boundaries for SAGB stars are not clear cut and depend on
different assumptions for convective overshooting
\citep{poelarends2008}.  The models used to construct the formulae used
by \citet{hurley2002} include overshooting and hence give SAGB stars
from initial masses $6.4-8.1\,\rm M_\odot$.  The fraction of SAGB
stars is then around $10^{-2}$ for a \citet{kroupa1993} initial
mass function.  The secondary must then be within a suitable mass
range for merging to take place.  This range is less restricted than
that for the separations.  For most of the systems that lead to a
$\gamma$-ray burst, the mass ratio $q$ is between 0.6 and 0.85. For a
flat distribution of mass ratio about 25\,per cent of the binary
systems fall within this range.  There are also suitable systems with
lower $q$ but the range of suitable separations for these is much
narrower.  Our binary population synthesis shows that the fraction of
binary systems with at least one component of initial mass above
$0.8\,\rm M_\odot$ that evolve to give a $\gamma$-ray burst is of the
order of $10^{-5}$.  Typically one such binary system is formed in our
own galaxy each year so this agrees well with the observed rarity of
gamma-ray bursts.

\begin{table}
 \centering
 \begin{minipage}{60mm}
   \caption{\label{table1}The fraction of binary systems $f$ with at
     least one component of mass $0.8\,\rm M_\odot$ that evolve to
     produce a $\gamma$-ray burst for various metallicities $Z$.  For
     our galaxy, at solar metallicity this fraction is numerically
     equal to the rate per year.
     }
  \begin{tabular}{@{}lc@{}}
  \hline
  $Z$ & $f$ \\
  \hline
  0.03 & $4.2\times 10^{-5}$ \\
  0.02 & $4.2\times 10^{-5}$ \\
  0.01 & $3.7\times 10^{-5}$ \\
  0.001 & $1.9\times 10^{-5}$ \\
  0.0001 & $2.5\times 10^{-5}$ \\
  \hline
\end{tabular}
\end{minipage}
\end{table}

Table~\ref{table1} lists the fractional rates for different
metallicity populations of binary stars.  As metallicity decreases our
estimated rate does not vary very much.  At a
metallicity of $Z=10^{-4}$ the frequency remains about the
same for the same star-formation rate and initial mass function.  The
suitable initial primary mass shifts to $5.1 < M_1/\rm M_\odot <6.8$
for an SAGB star in this lower metallicity environment, so the
frequency of suitable initial primary mass increases even if the IMF
is unchanged.  However, some of the low-$q$ systems can no longer
produce $\gamma$-ray bursts because the total mass of the system when
it merges is too low to trigger the accretion induced collapse.
Almost all the suitable systems have $q$ greater than 0.6.  These two
effects at lower metallicity balance each other out and hence
the the fraction of binary systems that lead to bursts remains of the
order of $10^{-5}$.  A different $q$ distribution which favoured the
high $q$ systems or a shift in the IMF towards intermediate-mass stars
would cause the frequency of $\gamma$-ray bursts to increase at low
metallicity.  Otherwise, our scenario shows that the $\gamma$-ray
burst rate does not have a high dependence of metallicity.  This would
explain the observational deduction of \citet{savaglio2010} that the
burst rate is proportional to the star formation rate alone.

\citet{fryer1999} raised concerns about the amount of neutron-rich
ejecta from accretion induced collapse events polluting the Galaxy.
Observed abundances of $r$-process isotopes place a rather low limit
on the rate of the events they model.
\citet{dessart2007} agree and place a limiting rate of about
$10^{-6}\rm\, yr^{-1}$ on their particular highly magnetic collapses, rising to
$5\times 10^{-5}\,\rm yr^{-1}$ for their less magnetic cases.  Given
the uncertainties in the models this is consistent with our GRB rate.
However we note that the accretion induced collapse events considered here,
because they take place within a hydrogen-free common envelope, are a
rather special subset of all such events.  Indeed the total rate
predicted in population synthesis calculations by \citet{hurley2002},
of at least $10^{-4}\,\rm yr^{-1}$ suggests that the amount of
neutron-rich material ejected by accretion induced collapse of white
dwarfs in general must be somewhat smaller than found by
\citet{fryer1999} and \citet{dessart2007}.

\section{Discussion}

In the scenario outlined above, the properties of the stellar cores
which eventually merge to give rise to a LGRB can differ from one
another in a number of respects.  First the amount of helium envelope
remaining at the time the cores merge can vary from several solar
masses to very little.  Secondly the mass of both cores can vary as
long as enough of the CO core can be accreted to drive the collapse of
the ONe core.  There is then a range from zero to about a solar
mass of CO that can accrete on to the neutron star.  Some of this can
burn and be ejected in a disc wind to provide the varying quantities
of $^{56}$Ni seen in the associated supernovae.  These range from the
very powerful, with as much as half a solar mass of nickel-56, to very
weak, with almost no nickel-56
\citep{mazzali2003,watson2007}.

We can also estimate the time required for the initial binary system
to evolve to the $\gamma$-ray burst.  At a metallicity of
$Z=10^{-4}$, the age of a typical system is around
$100-150\,$Myr. This gives the upper limit to the redshift of the
earliest gamma-ray burst of $z\approx 20$ and we should expect to see
bursts back to this redshift.  This is a prediction of our model.

Strongly magnetic neutron stars (magnetars)
have been identified in our galaxy as anomalous X-ray pulsars (AXPs)
or soft $\gamma$-ray repeaters \citep[SGRs,][]{mereghetti2008}.  These 
are observed to have magnetic fields
of $10^{14} - 10^{15}\,$G and spin periods of $2 - 10\,$s.  Known
magnetars have ages of $10^3 - 10^4\,$yr and it is unclear whether
they have spun down or were born slowly rotating.  With estimated
birthrates of $10^{-3}\,\rm yr^{-1}$ they are unlikely to all be
related to the LGRBs.  Magnetars could therefore be of two types,
those that are born from single star evolution and those that form
from the merging of the cores of two stars as proposed here.
As for the radio pulsars, the fields in the first group are
likely to be generated by a dynamo mechanism in the stellar core that
subsequently collapses to form the neutron star.  Strong magnetic
coupling with the stellar envelope leads to outward transport of
angular momentum during stellar evolution and results in a slowly
rotating neutron star following core collapse.  Magnetars that are
born by the merging of the cores of two stars are, on the other hand,
likely to be born rapidly spinning with high fields and so give rise
to LGRBs.

\section{Conclusions}

The $\gamma$-ray jets seen in the LGRBs have been attributed to MHD
extraction of the rotational energy of a neutron star or black hole,
or of a massive, about $0.1\,\rm M_\odot$, unstable disc that is
accreted by the compact star at the time of its birth.  The
combination of super strong magnetic fields and millisecond spin
periods that appear to be required to explain the energetics and
collimation of the $\gamma$-ray jet cannot easily be produced in
the stellar remnant through single star evolution. Nor does it appear
likely that single star evolution could lead to a rapidly spinning
disc around the compact star that could generate the field required to
produce the MHD jet.

We have argued that the required strong fields and rapid spins may
occur more naturally in binary star scenarios where merged stellar
cores collapse to form neutron stars or black holes and have presented
a convoluted but still simple binary star origin for $\gamma$-ray
bursts in which the progenitors are intermediate-mass stars and the
collapsed star is a strongly magnetic neutron star. This differs
radically from previous proposals that have envisaged
massive stars as the progenitors.

Our model predicts a birth rate of a few times $10^{-5}\,\rm yr^{-1}$
per galaxy at solar metallicity.  The fraction of binary stars that
lead to $\gamma$-ray bursts does not vary much with metallicity in our
model.  Thus we expect the rate of $\gamma$-ray bursts to be
proportional to the star formation rate for a wide range of redshifts.
We expect that the LGRBs should have characteristics of an
intermediate-mass population and need only be loosely linked to young
star forming regions.  Recent high resolution imaging of nearby LGRBs
appears to support this requirement.  The estimated time interval
between the time of formation of the component stars and the final
collapse to a neutron star is about $100\,$Myr so that LGRBs produced
by this channel should be seen up to red shifts of $z\approx 20$.

\section*{Acknowledgements}

CAT thanks Churchill College for his fellowship and the Australian
National University for generously supporting his visit to Australia
in 2009.  HHBL thanks the ANU for hospitality.

\label{lastpage}

\begin{thebibliography}{99}


\bibitem[\protect\citeauthoryear{Beckwith, Hawley \&
    Krolik}{2008}]{beckwith2008} Beckwith K., Hawley J. F., Krolik
  J. H., 2008, ApJ, 678, 1180 
\bibitem[\protect\citeauthoryear{Cantiello et
    al.}{2007}]{cantiello2007}Cantiello M., Yoon S.-C., Langer N.,
  Livio M., 2007, A\&A, 465, L29
\bibitem[\protect\citeauthoryear{Dessart et
al.}{2007}]{dessart2007}Dessart L., Burrows A., Livne E., Ott C. D.,
2007, ApJ, 669,585
\bibitem[\protect\citeauthoryear{Di Matteo, Perna \&
    Narayan}{2002}]{dimatteo2002}Di Matteo T., Perna R., Narayan R.,
  2002, ApJ, 579, 706
\bibitem[\protect\citeauthoryear{Duncan \& Thompson}
  {1992}]{duncan1992} Duncan R. C., Thompson C., 1992, ApJ, 392, L9
\bibitem[\protect\citeauthoryear{Eggleton, Fitchett \&
    Tout}{1989}]{eggleton1989}Eggleton P. P., Fitchett M. J., Tout
  C. A., 1989, ApJ, 347, 998
\bibitem[\protect\citeauthoryear{Fryer \&
    Heger}{2005}]{fryer2005}Fryer C. L., Heger A., 2005, ApJ, 623, 302
\bibitem[\protect\citeauthoryear{Fryer et al.}{1999}]{fryer1999}Fryer
C., Benz W., Herant M., Colgate S. A., 1999, ApJ, 516, 892
\bibitem[\protect\citeauthoryear{Fryer et al.}{2007}]{fryer2007}Fryer
  C. L., Mazzali P. A., Prochaska J. et al., 2007, PASP, 119, 1211
\bibitem[\protect\citeauthoryear{Giovannini et al.}{2001}]{giovannini2001}
  Giovannini G., Cotton W. D., Feretti L., Lara L.,
  Venturi T., 2001, ApJ, 552, 508
\bibitem[\protect\citeauthoryear{Giroletti et
    al.}{2004}]{giroletti2004} Giroletti M., Giovannini G.,Feretti L.,
  Cotton W. D., Edwards P. G., Lara L., Marscher A. P., Mattox J. R.,
  Piner B. G., Venturi T., 2004, ApJ, 600, 127
\bibitem[\protect\citeauthoryear{Hammer et al.}
  {2006}]{hammer2006}Hammer F., Flores H., Schaerer D.,
  Dessauges-Zavadsky M., Le Floc'h E., Puech M., 2006, A\&A, 454, 103
\bibitem[\protect\citeauthoryear{Heger, Woosley \& Spruit}
  {2005}]{heger2005} Heger A., Woosley  S. E., \& Spruit H. C., 2005,
  ApJ, 626, 350
\bibitem[\protect\citeauthoryear{Hough}{2008}]{hough2008}Hough D.,
  2008, in Rector T. A., De Young D. S., eds, ASP Conf. Ser. Vol. 386,
  Extragalactic Jets: Theory and Observation from Radio to Gamma Ray,
  Astron. Soc. Pac., San Francisco, p.~274
\bibitem[\protect\citeauthoryear{Hurley, Tout \&
    Pols}{2002}]{hurley2002}Hurley J. R., Tout C. A., Pols O. R., 2002, MNRAS,
  329, 897
\bibitem[\protect\citeauthoryear{Izzard, Ramirez-Ruiz \&
    Tout}{2004}]{izzard2004}Izzard R. G., Ramirez-Ruiz E., Tout C. A.,
  2004, MNRAS, 348, 1215
\bibitem[\protect\citeauthoryear{Klu\'zniak \&
    Ruderman}{1998}]{kluzniak1998}Klu\'zniak W., Ruderman M., 1998,
  ApJ, 505, L113
\bibitem[\protect\citeauthoryear{Komissarov \&
    Barkov}{2009}]{komissarov2009}Komissarov S. S., Barkov M. V., 2009,
  MNRAS, 397, 1153
\bibitem[\protect\citeauthoryear{Komissarov, Vlahakis \&
K\"onigl}{2010}]{komissarov2010}Komissarov S. S., Vlahakis N., K\"onigl
A., 2010, MNRAS, in press
  MNRAS, 397, 1153
\bibitem[\protect\citeauthoryear{Kroupa, Tout \&
    Gilmore}{1993}]{kroupa1993}Kroupa P., Tout C. A., Gilmore G.,
  1993, MNRAS, 262, 545
\bibitem[\protect\citeauthoryear{Lithwick \&
    Sari}{2001}]{lithwick2001}Lithwick Y., Sari R., 2001, ApJ, 555, 540
\bibitem[\protect\citeauthoryear{Livio, Ogilvie \&
    Pringle}{1999}]{livio1999}Livio M., Ogilvie G. I., Pringle J. E.,
  1999, ApJ, 512, 100
\bibitem[\protect\citeauthoryear{Lubow, Papaloizou \&
    Pringle}{1994}]{lubow1994}Lubow S. H., Papaloizou J. C. B.,
  Pringle J. E., 1994, MNRAS, 267, 235
\bibitem[\protect\citeauthoryear{Lyubarsky}{2009}]{lyubarsky2009}Lyubarsky
  Y., 2009, ApJ, 698, 1570
\bibitem[\protect\citeauthoryear{Lyubarsky}{2010}]{lyubarsky2010}Lyubarsky
  Y. E., 2010, MNRAS 402, 353
\bibitem[\protect\citeauthoryear{MacFadyen \&
    Woosley}{1999}]{macfadyen1999}MacFadyen A. I., Woosley S. E.,
  1999, ApJ, 524, 262
\bibitem[\protect\citeauthoryear{Mazzali et
    al.}{2003}]{mazzali2003}Mazzali P. A., Deng J., Tominaga N. et
  al., 2003, ApJ, 599, 95
\bibitem[\protect\citeauthoryear{Mereghetti}{2008}]{mereghetti2008}
  Mereghetti S., 2008, A\&A Rev., 15, 225
\bibitem[\protect\citeauthoryear{M\'esz\'aros \&
    Rees}{1997}]{meszaros1997}M\'esz\'aros P., Rees M. J., 1997, ApJ,
  476, 232
\bibitem[\protect\citeauthoryear{Paczy\'nski}{1976}]{paczynski1976}
Paczy\'nski B., 1976, in Eggleton P. P., Mitton S., Whelan J., eds,
Proc. IAU Symp. 73, Structure and Evolution of Close Binary Systems,
Reidel, Dordrecht, p.~75
\bibitem[\protect\citeauthoryear{Paczy\'nski}{1998}]{paczynski1998}Paczy\'nski
  B., 1998, ApJ, 494, L45
\bibitem[\protect\citeauthoryear{Podsiadlowksi et
    al.}{2004}]{podsiadlowski2004}Podsiadlowski P., Mazzali P. A.,
  Nomoto K., Lazzati D., Cappellaro E., 2004, ApJ, 607, L17
\bibitem[\protect\citeauthoryear{Poelarends et
    al.}{2008}]{poelarends2008} Poelarends A. J. T., Herwig F., Langer
  N., Heger A., 2008, 675, 614
\bibitem[\protect\citeauthoryear{Porth \&
Fendt}{2010}]{porth2010}Porth O., Fendt C., 2010, ApJ, 709, 1100
\bibitem[\protect\citeauthoryear{Price, Pringle \&
    King}{2003}]{price2003}Price D. J., Pringle J. E., King A. R.,
  2003, MNRAS, 339, 1223 
\bibitem[\protect\citeauthoryear{Salvaterra et
    al.}{2009}]{salvaterra2009}Salvaterra R., Della Valle M., Campana
  S. et al., 2009, Nat, 461, 1258
\bibitem[\protect\citeauthoryear{Savaglio}{2010}]{savaglio2010}Savaglio
  S., 2010, in Cunha K., Spite M., Barbuy B., eds, Proc. IAU
Symp.~265, Chemical Abundances in the Universe: Connecting First Stars
to Planets. Cambridge Univ. Press, Cambridge, p.~139
\bibitem[\protect\citeauthoryear{Sollerman et
    al.}{2005}]{sollerman2005}Sollerman J., \"Ostlin G., Fynbo
  J. P. U., Hjorth J., Fruchter A., Pedersen K., 2005, New Astron.,
  11, 103
\bibitem[\protect\citeauthoryear{Spruit}{1999}]{spruit1999}Spruit
  H. C., 1999, A\&A, 341, L1
\bibitem[\protect\citeauthoryear{Spruit}{2002}]{spruit2002}Spruit
  H. C., 2002, A\&A, 381, 923
\bibitem[\protect\citeauthoryear{Tchekhovskoy, McKinney \&
    Narayan}{2008}]{tchekhovskoy2008}Tchekhovskoy A., McKinney J. C.,
  Narayan R., 2008, MNRAS, 388, 551
\bibitem[\protect\citeauthoryear{Tout et al.}{2008}]{tout2008}Tout
  C. A., Wickramasinghe D. T., Liebert J., Ferrario L., Pringle J. E.,
2008, MNRAS, 387, 897
\bibitem[\protect\citeauthoryear{Uomoto}{1986}]{uomoto86}Uomoto A., 1986,
  ApJ, 310, L35
\bibitem[\protect\citeauthoryear{Usov}{1992}]{usov1992}Usov V. V., 1992,
  Nat, 357, 472
\bibitem[\protect\citeauthoryear{Watson et
    al.}{2007}]{watson2007}Watson D., Fynbo J. P. U., Th\"one C. C.,
  Sollerman J., 2007, Phil. Trans. R. Soc. A, 365, 1269
\bibitem[\protect\citeauthoryear{Wheeler \&
    Levreault}{1985}]{wheeler85}Wheeler J. C., Levreault R., 1985, ApJ, 294, L17
\bibitem[\protect\citeauthoryear{Woosley}{1993}]{woosley1993}Woosley
  S. E., 1993, ApJ, 405, 273
\bibitem[\protect\citeauthoryear{Woosley \&
    Bloom}{2006}]{woosley2006a}Woosley S. E., Bloom J. S., 2006,
  ARA\&A, 44, 507
\bibitem[\protect\citeauthoryear{Woosley \&
    Heger}{2006}]{woosley2006}Woosley S. E., Heger A., 2006, ApJ, 637, 914
\bibitem[\protect\citeauthoryear{Yi \& Blackman}{1997}]{yi1997}Yi I.,
Blackman E. G., 1997, ApJ, 482, 383
\bibitem[\protect\citeauthoryear{Yoon \& Langer}{2005}]{yoon2005}Yoon
  S.-C., Langer N., 2005, A\&A, 443, 643
\bibitem[\protect\citeauthoryear{}{}]{}

\end{thebibliography}
\end{document}